\documentclass[conference]{IEEEtran}
\IEEEoverridecommandlockouts
\usepackage{authblk}
\usepackage{cite}
\usepackage{amsmath,amssymb,amsfonts}
\usepackage{amsmath}  
\usepackage{algorithm}  
\usepackage{algorithmicx}  
\usepackage{algpseudocode}  
\usepackage{amsfonts}  
\usepackage{graphicx}
\usepackage{textcomp}
\usepackage[table]{xcolor}

\usepackage{tabularx}
\usepackage{url}
\usepackage[bookmarks=false]{hyperref}
\usepackage{rotating}
\usepackage{subcaption}
\usepackage{multirow}
\usepackage{pifont}
\usepackage{tcolorbox}
\definecolor{tumColorLightBlue}{HTML}{f0f5fa} 
\usepackage{cleveref}
\usepackage{makecell}

\usepackage{tablefootnote}
\usepackage{threeparttable}
\usepackage{multirow}
\usepackage{booktabs}
\usepackage{harveyballs}

\renewcommand{\baselinestretch}{0.92}

\usepackage{eso-pic}
\usepackage[outdir=./]{epstopdf}
\usepackage{fancyhdr}

\def\BibTeX{{\rm B\kern-.05em{\sc i\kern-.025em b}\kern-.08em
    T\kern-.1667em\lower.7ex\hbox{E}\kern-.125emX}}

\AddToShipoutPictureBG*{%
  \AtPageUpperLeft{%
    \put(0.07\paperwidth,-1.2cm){%
      \fbox{%
        \begin{minipage}{\textwidth}
          \centering
          \small
          \textbf{This paper has been accepted for publication in the proceedings of the 2026 IEEE 104th Vehicular Technology Conference (VTC2026-Fall), September 2026. The copyright for this paper will be transferred to IEEE. ©2023 IEEE. Personal use of this material is permitted. Permission from IEEE must be obtained for all other uses, in any current or future media, including reprinting/republishing this material for advertising or promotional purposes, creating new collective works, for resale or redistribution to servers or lists, or reuse of any copyrighted component of this work in other works.}
        \end{minipage}
      }%
    }%
  }%
}

\begin{document}
\title{Introducing Large Language Models into the Design Flow of Time-Sensitive Networking}

\author[1]{Rubi Debnath}
\author[2]{Luxi Zhao}
\author[3]{Mohammadreza Barzegaran}
\author[4]{Paul Pop}
\author[5]{Sebastian Steinhorst\vspace{-0.2cm}} 

\affil[1,5]{TUM School of Computation, Information and Technology, Technical University of Munich, Germany}
\affil[2]{Electronic Information Engineering College, Beihang University, Beijing, China}
\affil[3]{Department of Electrical Engineering and Computer Science, University of California, Irvine, USA}
\affil[4]{DTU Compute, Technical University of Denmark, Denmark}
{
    \makeatletter
    \renewcommand\AB@affilsepx{, \protect\Affilfont}
    \makeatother
    \affil[1,5]{firstname.lastname@tum.de}
    \affil[2]{zhaoluxi@buaa.edu.cn}
    \affil[3]{barzegm1@uci.edu}
    \affil[4]{paupo@dtu.dk}
}
\makeatletter
\patchcmd{\@maketitle}
  {\addvspace{0.5\baselineskip}\egroup}
  {\addvspace{-1\baselineskip}\egroup}
  {}
  {}
\makeatother

\maketitle

\begin{abstract}
The growing demand for real-time, safety-critical systems has significantly increased both the adoption and complexity of Time-Sensitive Networking (TSN). Configuring an optimized TSN network is highly challenging, requiring careful planning, design, analysis, verification, validation, and deployment. Large Language Models (LLMs) have recently demonstrated strong capabilities in solving complex tasks, positioning them as promising candidates for automating end-to-end TSN deployment and management, referred to as TSN orchestration. This paper outlines the steps involved in TSN orchestration and the associated challenges. To assess the capabilities of existing LLMs, we conduct an initial proof-of-concept case study focused on TSN tasks across multiple models. Building on these insights, we propose an LLM-assisted orchestration framework. Unlike prior research on LLMs in computer networks, which has concentrated on general configuration and management, TSN-specific orchestration has not yet been investigated. We present the building blocks for automating TSN using LLMs, describe the proposed pipeline, and analyze opportunities and limitations for real-world deployment. This work provides the first roadmap toward assessing the feasibility of LLM-assisted TSN orchestration. 
\end{abstract}

\vspace{-0.3cm}
\section{Introduction}
\label{sec:introduction}
\begin{figure}[!t]
    \centering
    \includegraphics[scale=0.46, trim={0.6cm 0.5cm 0.6cm 0.4cm}, clip]{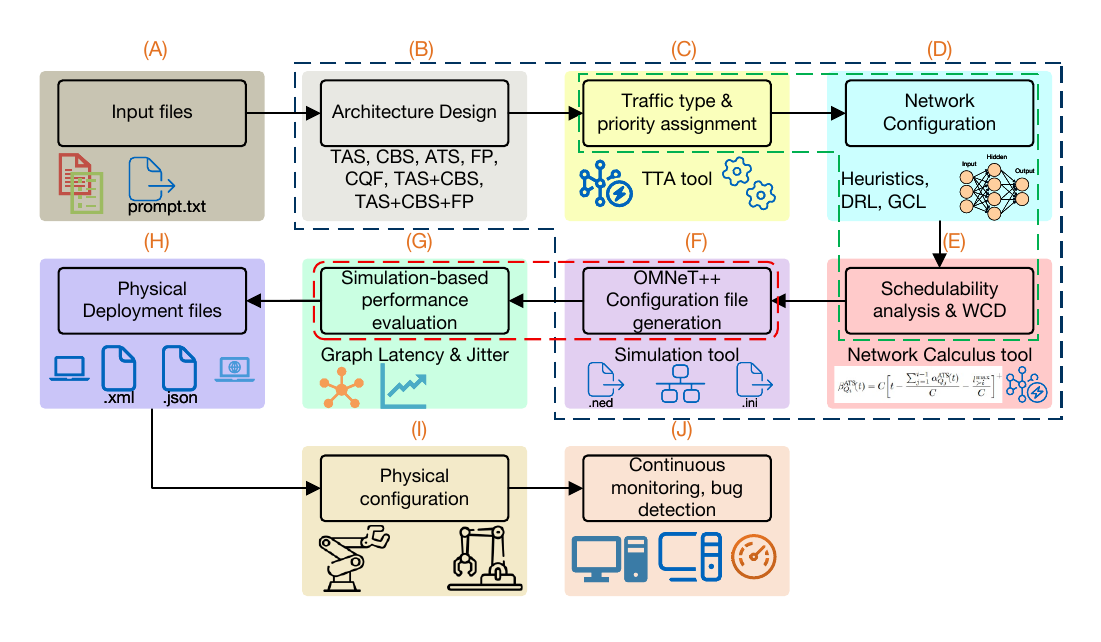}
    \caption{Proposed LLM-aided TSN Orchestrator architecture. The pipeline illustrates the key steps in the end-to-end orchestration and deployment of TSN networks.}
    \label{fig:fig1}
    \vspace{-0.4cm}
\end{figure}

Time-Sensitive Networking (TSN)~\cite{gcl_silviu} provides deterministic communication and bounded latency, making it essential for safety-critical domains such as industrial automation, vehicular systems, aerospace, healthcare, and emerging 5G-TSN applications. By supporting mixed-criticality traffic~\cite{voica_traffic_assignment} on the same network, TSN enables diverse Quality of Service (QoS) requirements to coexist. However, configuring an optimized TSN network remains highly challenging. It involves multiple building blocks (refer to Fig.~\ref{fig:fig1}) including planning, design, verification, validation, and deployment, all of which demand deep expertise in TSN. This process is time-consuming, tedious, and heavily dependent on domain experts.

Meanwhile, Large Language Models (LLMs) have achieved rapid adoption across research and industry, demonstrating impressive capabilities in solving complex, multi-step tasks. Prior studies have demonstrated their effectiveness for mathematical reasoning~\cite{realmath}, network configuration~\cite{llm_zero_touch, netconfeval_conext_2024}, network planning~\cite{ChatNet}, and fifth-generation (5G) network analysis~\cite{5g_llm}. In industry, LLMs are already being explored for automating infrastructure management and operational support. Despite this progress, their application to TSN remains largely unexplored. Unlike general computer networking, TSN introduces unique challenges, including deterministic guarantees, network configuration~\cite{gcl_silviu, rubi_iotj}, schedulability analysis~\cite{luxi_tnsm}, traffic type assignment (TTA)~\cite{voica_traffic_assignment, rubi_icc}, simulative evaluation~\cite{rubi_rtcsa, rubi_noms} and a lack of open-source tools, datasets and sub-standards. TSN is often used in safety-critical systems, so the outputs of an LLM-assisted orchestrator cannot be judged only by syntactic correctness or convenience. In safety assurance~\cite{graydon2013safety}, a workflow must provide the properties and evidence needed to justify timing and deployment decisions.

\subsection{Motivation and Contributions}
\label{sec:motivation}
Due to the inherent complexity of TSN and the high level of domain expertise required for deployment, there is a need for alternative approaches that can streamline the end-to-end process. Considering the huge capabilities of LLMs, our goal is to explore their potential for automating key stages of real-world TSN deployment. In this paper, we use the term \textbf{TSN orchestration} to refer to the end-to-end pipeline required to configure, validate, and deploy a TSN network (step A to J of Fig.~\ref{fig:fig1}). An LLM may participate in some or all of these steps (shown in Fig.~\ref{fig:fig1}), depending on the level of automation desired. The ability of LLM models to process natural language instructions, model configurations, and iteratively refine solutions makes them attractive for handling multi-stage orchestration pipelines. However, no prior work has investigated their use in TSN-specific orchestration. This gap motivates our work to perform the initial evaluation of LLMs with TSN tasks. Our main contributions are:

\begin{itemize}
    \item We introduce a conceptual framework illustrating how LLMs can support planning, design, verification, validation, and deployment of TSN networks. We identify orchestration stages (refer to Fig.~\ref{fig:fig1}) where LLMs can be effective and highlight those that require hybrid integration with external tools.  
    \item We present an illustrative case study across several state-of-the-art models to showcase current capabilities and limitations of the models. 
    \item Lastly, we discuss the challenges, limitations, and future research directions of LLMs in TSN.
\end{itemize}

Overall, this paper serves as a starting point for the community to explore LLM-assisted orchestration of TSN. 

\section{Background}
\label{sec:background}
TSN is a layer-2 mechanism that provides deterministic communication and bounded latency. TSN achieves this by utilizing shaping and scheduling mechanisms, including the Time Aware Shaper (TAS)~\cite{rubi_noms}, Credit Based Shaper (CBS)~\cite{rubi_noms}, Asynchronous Traffic Shaper (ATS)~\cite{luxi_tnsm}, Cyclic Queuing and Forwarding (CQF)~\cite{rubi_ccnc}, and Frame Preemption (FP)~\cite{rubi_noms}. 

Furthermore, for the configuration and scheduling of TSN flows in the network, optimized algorithms are required. These algorithms are often developed using mathematical models~\cite{gcl_silviu}, heuristics~\cite{rubi_iotj}, and advanced machine learning models~\cite{rubi_icc}. However, developing optimal scheduling and configuration algorithms is NP-hard~\cite{rubi_iotj}, which requires domain-specific algorithms to reduce time complexity.

The schedulability analysis further provides the mathematical upper bounds of the worst-case delay (WCD) of the TSN flows in the network. For this, Network Calculus (NC) tools~\cite{luxi_tnsm} are widely used both in academia and industry. This process is widely known as schedulability analysis or WCD calculation. These tools compute upper bounds on delay and jitter, which is particularly crucial for event-triggered (ET) traffic, especially when assessing the impact of interference from co-existing time-triggered (TT) flows.

Since the worst-case bounds derived from NC can be pessimistic, simulation frameworks such as OMNeT++~\cite{rubi_noms, rubi_ccnc} are often used to complement the analysis. These simulations help assess the system's performance under statistical conditions, providing insights into the practical gap between the theoretical worst-case and observed behavior. 

Despite their importance, tools such as NC models, TSN-specific scheduling and configuration algorithms are rarely open source, limiting broader adoption and automation. Furthermore, all these steps are typically evaluated and optimized individually. Each of these models are inherently complex, requiring domain experts to carefully develop specific algorithms and tools. 

\subsection{LLMs in Networking}
LLMs have recently demonstrated strong capabilities in solving complex reasoning and multi-step tasks. They excel at parsing natural language, synthesizing structured outputs, and integrating information across diverse sources, making them promising for networking tasks. Recent works~\cite{5g_llm, ChatNet, netconfeval_conext_2024} have explored the use of LLMs for various computer networking problems. For example, \cite{5g_llm} introduced Mobile-LLaMA, an open-source, mobile-network-specialized LLM designed for 5G network management and data analysis. Huang et al. in \cite{ChatNet} proposed ChatNet, a domain-adapted LLM framework that reduces the complexity of network planning tasks. In \cite{netconfeval_conext_2024}, the authors demonstrated how LLMs can generate protocol-specific configurations by leveraging newly acquired knowledge from the internet. Other works~\cite{llm_zero_touch, netconfeval_conext_2024} have demonstrated LLM-driven configuration synthesis, protocol-specific configuration generation, and knowledge acquisition from online sources. 

\textit{Existing TSN orchestration relies heavily on domain-specific algorithms, optimization tools, and custom scripts. These approaches are rigid, and require significant manual engineering. LLMs introduce a new leap by offering a natural-language interface that allows even non-experts to express requirements in plain English (for example, “set up a TSN network with three TSN flows with 1 ms deadlines using TAS”).} 

\subsection{Research Questions}
While prior research has successfully applied LLMs to general networking problems, no existing work has addressed their use for TSN-specific orchestration. TSN introduces additional challenges such as deterministic guarantees, schedulability analysis, and mixed-criticality flow assignment, which remain unexplored in the context of LLMs. This gap motivates our work to outline the steps, workflow, and challenges of developing LLM-assisted TSN orchestration. In this paper, we aim to answer the following research questions:

\vspace{0.2cm}
\noindent \textit{\textbf{RQ 1:} Can LLMs help with orchestrating TSN networks?} \\
\noindent \textit{\textbf{RQ 2:} What steps are required for LLMs to effectively orchestrate TSN networks?} \\
\noindent \textit{\textbf{RQ 3:} What are the challenges and future directions of LLM-assisted TSN orchestration?}
\vspace{0.2cm}

As shown in Fig.~\ref{fig:fig1}, multiple steps are involved in progressing from initial planning to real-world TSN deployment. We separate these steps and map them into LLM-assisted building blocks. The proposed LLM-assisted orchestration workflow consists of: (A) understanding topology, flow information, and network parameters, (B) architecture and mechanism selection, (C) traffic type and priority assignment, (D) network configuration or schedule synthesis, (E) schedulability analysis and WCD calculation, (F) OMNeT++ simulation files, (G) simulation-based performance evaluation and verification, (H) physical deployment files, (I) physical configuration, and (J) continuous monitoring and bug detection (shown in Fig.~\ref{fig:fig1}). 

\section{End-to-End TSN Orchestration using LLMs}
\label{sec:llm_tsn}
Unlike general network configuration steps described in~\cite{netconfeval_conext_2024}, TSN orchestration requires several TSN-specific stages as shown in Fig.~\ref{fig:fig1}. As discussed, related work focuses on each section of Fig.~\ref{fig:fig1} individually. Therefore, there is currently a lack of research addressing the entire pipeline together. An efficient LLM-based model for TSN should execute a sequence of well-defined tasks to achieve complete end-to-end orchestration. We present a roadmap of these tasks, with the proposed pipeline illustrated in Fig.~\ref{fig:fig1}. These tasks have different assurance implications. Input parsing, deployment file generation, and monitoring summaries mainly support engineering productivity. In contrast, traffic assignment, schedule synthesis, schedulability analysis, and physical deployment influence whether timing and safety requirements are satisfied. For this latter group, a one-shot prompt-answer workflow may be insufficient with the current LLM capabilities. Instead, the LLMs should be used as a translation and orchestration layer around solvers, NC engines, simulators, validators, and, where relevant, model checkers or theorem provers, while these tools remain the source of the evidence used in the final decision. 

\subsection{Input Parsing}
Users may provide either (a) a natural-language description of the topology and flows or (b) structured files containing flow and topology parameters. While natural language is feasible for small scenarios, structured input is more practical for large-scale networks with hundreds of flows. The LLM extracts information from these files and converts it into a machine-readable format. As noted in~\cite{llm_workflow_network}, describing flow properties purely in natural language can lead to information loss. We therefore use a unified structure for describing flows, which the LLM processes for further steps.

\subsection{Architecture Design}
TSN supports multiple combinations of shapers and schedulers (e.g., TAS, CBS, CQF, FP). The architecture may be predefined by the system engineer or suggested by the LLM based on QoS requirements. The model iterates over candidate designs, refining choices using validation feedback. At this stage, the LLM determines the mechanisms or TSN architecture to be deployed and explores alternative designs until performance targets are satisfied.

\subsection{Traffic Type and Priority Assignment}
The model maps flows to TSN traffic types and priorities. Given a set of flows, if the TSN traffic type and priority are not known or predefined in the user prompt, the model examines the requirements and description of the prompt and, based on domain knowledge, assigns suitable traffic types and priorities. An efficient model should iteratively refine the traffic type and priority assignment through schedulability and simulation-based performance analysis. \textit{Since TTA is NP-hard, external solvers such as~\cite{rubi_icc} may be integrated to enhance efficiency and accuracy. Furthermore, the architecture given by the LLM should be verified using external solvers.} LLM models are unlikely to achieve optimal accuracy without incorporating external TTA tools.

\subsection{Network Configuration}
TAS works based on the predefined static table known as the Gate Control List (GCL), also denoted as the scheduling table. CQF and other cyclic shaper variants require the network configuration and the offset (start time of the flow in the source node) values. These are determined by specific algorithms designed to solve these NP-hard problems. Based on the architecture and the traffic type assigned by the model, the scheduling and configuration mechanisms are derived in this stage. Traditional algorithms are developed using heuristics, metaheuristics, and machine learning models. \textit{LLMs could integrate existing state-of-the-art optimization algorithms for scheduling and configuration. In this hybrid approach, the LLM will automate the algorithm selection (from a bunch of existing algorithms), input preparation, and iterative refinement. This enables to deliver near-optimal solutions with reduced manual effort and lower operational complexity, while maintaining the mathematical guarantees provided by the underlying optimization tools.} 

\subsection{Schedulability Analysis}
Schedulability analysis verifies whether the flows meet the QoS requirements (such as deadlines) by computing the WCD using NC tools~\cite{luxi_tnsm}. To perform the schedulability analysis, the network configuration from the previous step is required. \textit{As LLMs are not reliable for precise mathematical reasoning~\cite{ChatNet}, external NC modules are required to compute accurate delay bounds. The LLM prepares the NC inputs by translating natural-language or structured descriptions of flows, priorities, and schedules into NC-compatible formats (such as constraints and equations). The output generated by the scheduling algorithm may not be compatible with the NC model input, and LLM can solve this by adapting the format of the files. It can further interpret and summarize the NC outputs, making them understandable to non-experts.} Furthermore, the LLM feeds the results back into the traffic type and priority configuration for further refinement, as discussed in~\cite{rubi_icc}. 

\subsection{OMNeT++ Configuration File Generation}
Schedulability analysis provides the NC-based WCD analysis of the TSN flows in the network, which informs the theoretical upper bounds of delay and jitter. However, to achieve a complete performance evaluation of the network, simulation-based evaluation is also required. In traditional approaches, OMNeT++ based performance evaluations can be conducted in parallel with schedulability analysis. We need the network configuration, traffic types, and the priorities of the flows to perform simulation-based evaluation. OMNeT++ configuration files are typically generated using custom Python scripts or manually written by simulation developers. Creating OMNeT++ configuration files manually for very large and complex networks is tedious and error-prone. \textit{Therefore, in this stage, the LLM model generates the OMNeT++ configuration files for validation and performance evaluation. More specifically, we are interested in the \textit{.ned} and \textit{.ini} files of OMNeT++, which the model should provide without manual intervention. These files should be generated with the appropriate syntax and structure, which can be learned by the model during training. The model can also perform self-checks to evaluate syntax errors in the generated OMNeT++ files.} 

\subsection{Simulation-based Performance Evaluation}
Performance evaluation verifies and validates the generated architecture and configurations. In this stage, we refer to simulation-based evaluation using state-of-the-art simulators and frameworks~\cite{rubi_ccnc, rubi_noms}. While syntax errors are addressed earlier, this stage focuses on assessing network performance using simulation frameworks such as OMNeT++ and NS-3. The LLM produces simulation-ready configuration files (e.g., \texttt{.ned}, \texttt{.ini}) in the previous stage, and in this stage we execute the simulations. The simulation can be performed on an external OMNeT++ framework. 

\subsection{Physical Deployment Files}
Following successful verification and validation of the network, we propose that the LLM produce deployment-ready network configuration files, including routes, VLAN tags, and other TSN-specific configuration files. These deployment files are intended for real-world industrial deployment and are therefore distinct from the OMNeT++ configuration files generated in the previous stage. Prior research has shown that LLMs can generate router configurations~\cite{llm_router} efficiently. Consequently, fine-tuned models can be adapted to produce TSN-specific deployment files. \textit{Furthermore, the models can be refined to provide vendor-specific (such as Cisco, Hirschmann, NXP) templates and configuration files. By training on vendor manuals and sample configurations, LLMs can generate vendor-specific deployment files, ensuring compatibility across different platforms.}

\subsection{Physical Configuration}
The deployment files are then applied to the industrial or application-specific infrastructure. \textit{At this stage, zero-touch configuration~\cite{llm_zero_touch} can be adopted to provision and configure the entire network with a single operation, minimizing manual intervention and reducing the potential for human error.}

\subsection{Continuous Monitoring and Bug Detection}
Continuous monitoring, log analysis, and debugging are crucial to detect errors, bottlenecks, and failures. \textit{LLMs can process logs in real time, identify issues, trigger alarms, and suggest corrective actions. In dynamic environments where new TSN flows are added or existing ones removed, the LLM assists in runtime reconfiguration.}

In summary, this paper provides a sequential orchestration workflow in which each block builds on the previous one, while results from the schedulability and simulation stages feed back into earlier decisions for refinement and improvement. This iterative design ensures compliance with QoS requirements while reducing manual effort and increasing scalability.

\begin{figure}[!t]
    \centering
    \includegraphics[scale=0.5, trim={0.6cm 0.6cm 0.6cm 1.7cm}, clip]{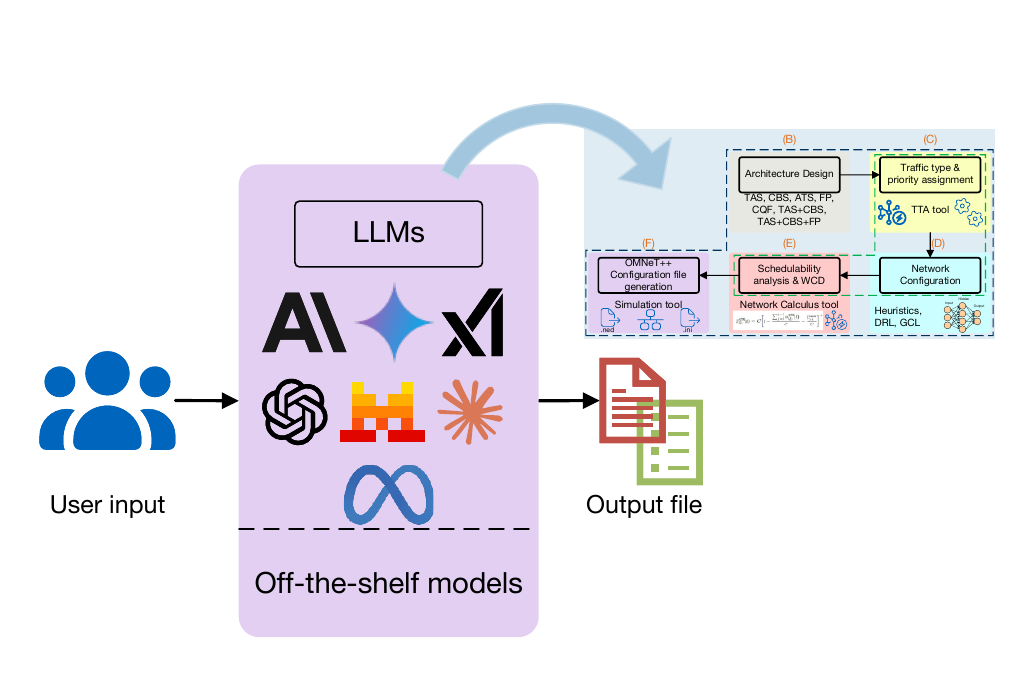}
    \caption{Overview of the case study setup.} 
    \label{fig:fig2}
    \vspace{-0.4cm}
\end{figure}

\section{Initial Case Study: Proof-of-Concept}
\label{sec:case_study}
To illustrate the potential of LLMs for TSN orchestration, we conducted an initial proof-of-concept study using several state-of-the-art models. In this study, we evaluate the models on orchestration steps (B–F) (refer to Fig.~\ref{fig:fig2}). We only cover steps B--F to narrow the focus of the framework, making it easier to provide readers with an intuitive understanding of how current LLMs behave when tasked with TSN-related orchestration. The prompt was designed to cover orchestration steps (B), (C), (D), (E), and (F) in Fig.~\ref{fig:fig1}. This section provides an illustrative example that complements the overall vision presented in Fig.~\ref{fig:fig1}. We consider four progressively complex test cases: (1) three end stations connected via two switches (refer to Fig.~\ref{fig:topo_one}) with three flows, and Erdos Renyi Graph (ERG)~\cite{rubi_ccnc} topologies (refer to Fig.~\ref{fig:topo_erg}) with (2) five flows, (3) ten flows, and (4) twenty flows. All experiments were conducted on a Windows laptop equipped with an Intel\textregistered\ Processor Core\texttrademark\ i7-10610U CPU running at 1.80GHz, and 32 GB RAM. Each case study is evaluated using selected state-of-the-art models from Table~\ref{tab:llm_benchmarking_matrix}. For consistency, all models are prompted with identical input descriptions, and their responses are compared in terms of OMNeT++ configuration correctness, completeness of outputs, and traffic type identification. We verify the syntax correctness and completeness of the OMNeT++ simulation files. 
\begin{table*}[t]
\centering
\renewcommand{\arraystretch}{1.2}
\begin{tabular}{|c|c|c|c|c|c|c|c|}
\hline
\textbf{Model} &
\textbf{Company} &
\makecell{\textbf{Access}\\\textbf{Mode}} &
\makecell{\textbf{Token}\\\textbf{Usage}} &
\makecell{\textbf{Cost}\\\textbf{(USD)}} &
\makecell{\textbf{OMNeT++ Config}\\\textbf{Accuracy}} &
\makecell{\textbf{Output}\\\textbf{Completeness}} &
\makecell{\textbf{Traffic Type}\\\textbf{Identification}} \\
\hline
GPT-5 & OpenAI & API & \centering 77699 & \$ 0.742 & \centering \makecell{Partially Correct} & \centering \makecell{Partially Complete} & TT \\
\hline
GPT-4o & OpenAI & API & \centering 8419 & \$ 0.054 & \centering Incorrect & \centering Incomplete & \makecell{Not Identified} \\
\hline
GPT-4o-mini & OpenAI & API & \centering 9290 & \$ 0.004 & \centering Incorrect & \centering \makecell{Partially Complete} & \makecell{Not Identified}\\
\hline
GPT-o3 & OpenAI & API & \centering 24042 & \$ 0.168 & \centering Incorrect & \centering \makecell{Partially Complete} & \makecell{TT, AVB Class A, AVB Class B}\\
\hline
GPT-o1-mini & OpenAI & API & \centering 18966 & \$ 0.069 & \centering Incorrect & \centering \makecell{Partially Complete} & \makecell{Not Identified}\\
\hline
Claude 3.7 & Anthropic & API & \centering 9361 & \$ 0.428 & \centering \makecell{Not Attempted} & \centering Incomplete & \makecell{Not Identified} \\
\hline
\makecell{Gemini 2.5 Pro\\(gemini-2.5-pro)} & \makecell{Google\\DeepMind} & API & \centering 25115 & \$ 0.211 & \centering\makecell{Partially Correct}& \centering \makecell{Partially Complete} & TT \\
\hline
Grok 4 & xAI & API & \centering 20568 & \$ 0.229 & \centering Incorrect & \centering Incomplete & TT \\
\hline
Mistral Large & Mistral AI & API & \centering 17054 & \$ 0.083 & \centering Incorrect & \centering Partially Complete & \makecell{TT, AVB Class A, AVB Class B}\\
\hline
LLaMA 3.1-70B & Meta & \makecell{Self Host\\ (Ollama)} & \centering NA & NA & \centering Incorrect & \centering Incomplete & \makecell{Priorities without Traffic Types}\\
\hline
\multicolumn{8}{|p{0.95\linewidth}|}{
\footnotesize
\textbf{OMNeT++ Config Accuracy:} \textit{Partially Correct}: Model generated both \texttt{.ned} and \texttt{.ini} files with syntax/parameter errors. \textit{Incorrect}: Files missing, incomplete, hallucinated, or contained major invalid syntax. \textit{Not Attempted}: Model did not try to generate simulation files.

\textbf{Output Completeness:} \textit{Complete}: Full answer provided with all required fields. \textit{Partially Complete}: Some fields missing or truncated. \textit{Incomplete}: Major parts missing or answer cut off.

\textbf{Traffic Type Identification:} Whether the model explicitly assigned flows to TSN traffic types (TT, AVB Class A, AVB Class B, etc.). \textit{Not Identified} means no traffic types are assigned to the flows. \textit{Priorities without Traffic Types} means the priorities are assigned to the flows but the traffic types are not mentioned.
} \\
\hline
\end{tabular}
\caption{Illustrative case study for TSN orchestration, highlighting token usage, cost, and output quality without fine-tuning.}
\label{tab:llm_benchmarking_matrix}
\vspace{-0.2cm}
\end{table*}

\vspace{-0.4cm}
\subsection{Prompt}
We present a sample prompt used for the experiments. The files \textit{topology.txt} and \textit{flows.txt} contain the network topology and flow information, respectively, provided in a structured format.
\label{sub:prompt}
\vspace{-0.2cm}
\begin{tcolorbox}[colback=gray!5!white,
colframe=black!75,
title={Example Prompt used in the case study}, fonttitle=\bfseries]
\small   
You are a TSN orchestrator. Suggest the TSN architecture, traffic type, and traffic priority for this network.\\
\textbf{Topology:} \textit{topology.txt} 
Links: 1 Gbps, 5 $\mu$s propagation delay, cut-through switches. \\
\textbf{Flows:} \textit{flows.txt}\\
\textbf{Constraints:}  \\
Ensure deadlines are satisfied. Ensure Quality of Service is met. The topology is given as text. Switch represents the TSN switch. PLC represents the end stations. WIRE represents the connection. Flow information is given as text file.\\ 
\textbf{Output:} \\ 
Provide the architecture of the TSN network and mention which mechanisms to use. Mention the flow priority. Mention the traffic type of the flow. Perform the scheduling, configuration, and schedulability analysis using mathematical models. Find the worst case delays and jitter. For the generated architecture give the OMNeT++ simulation file for verification. Provide the output in text.
\end{tcolorbox}

\subsection{Outcomes}
Several insights emerge from this experiment and are listed as follows:

\begin{figure}[!t]{}
\begin{minipage}[b]{0.23\textwidth}
\centering
\includegraphics[scale=0.26, trim={0cm 0cm 0cm 0cm}, clip]{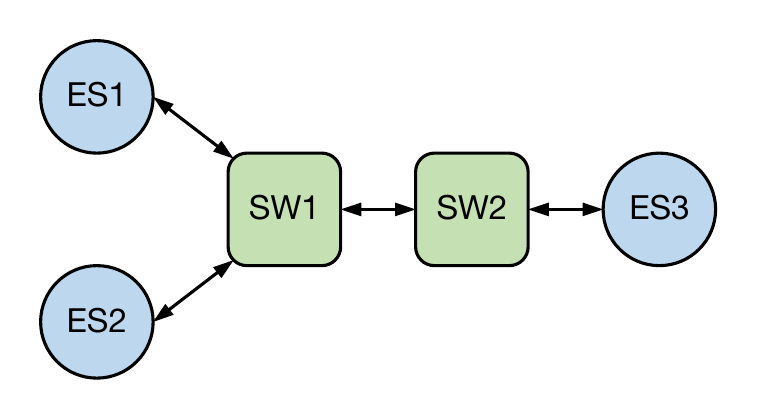}
\subcaption[]{Simple}
\label{fig:topo_one}
\end{minipage}
\hfill
\begin{minipage}[b]{0.23\textwidth}
\centering
\includegraphics[scale=0.26, trim={0cm 0cm 0cm 0cm}, clip]{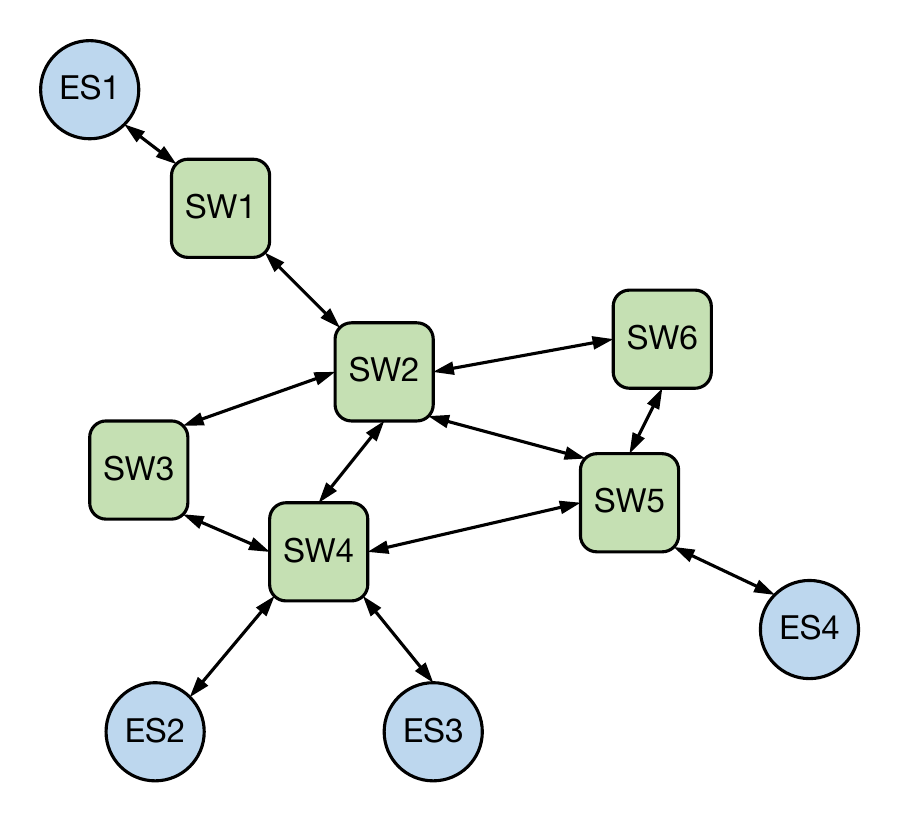}
\subcaption[]{ERG}
\label{fig:topo_erg}
\end{minipage}
\caption{Test Topologies}
\vspace{-0.4cm}
\end{figure}

\begin{figure}[!t]
    \centering
    \includegraphics[scale=0.44, trim={0.4cm 0.4cm 0.4cm 0.4cm}, clip]{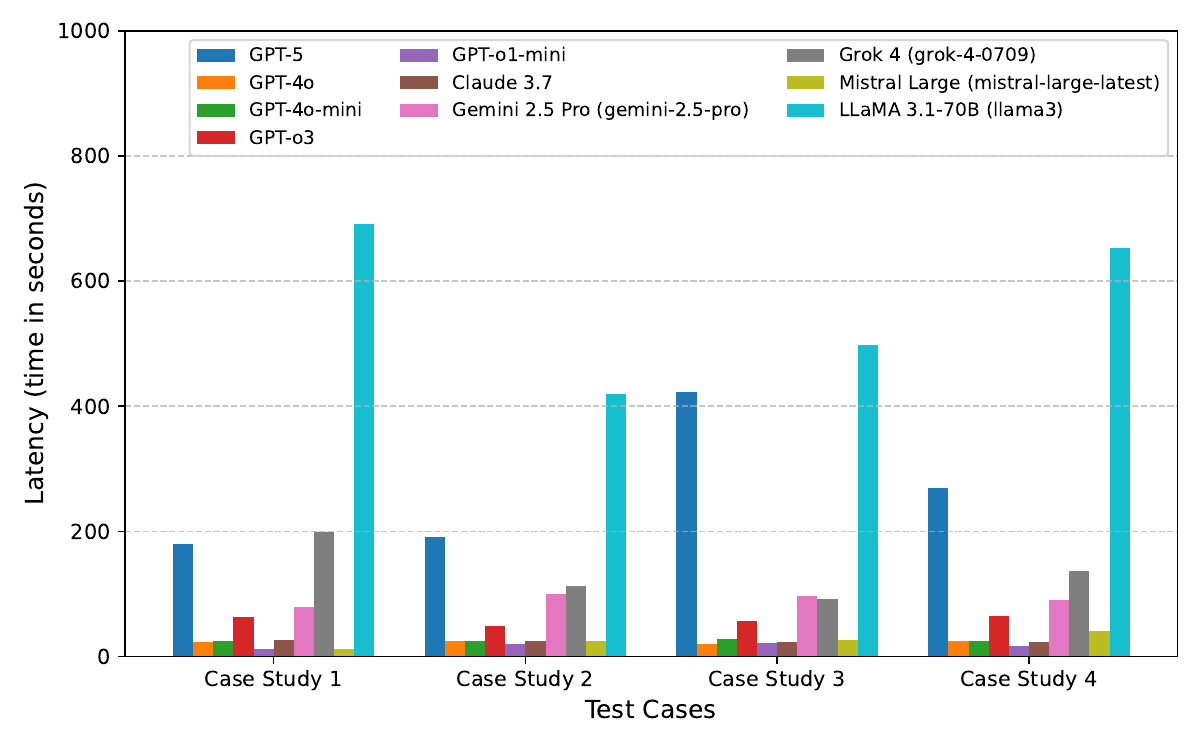}
    \caption{Latency across different LLM models.}
    \label{fig:llm_latency}
    \vspace{-0.3cm}
\end{figure}

\begin{itemize}
    \item \textbf{Latency:} Latency is an important parameter when using LLMs. API-based models such as GPT-5, GPT-4o, Claude 3.7, Gemini 2.5 Pro, Grok 4, and Mistral Large returned results with relatively low latency. In contrast, the LLaMA 3.1-70B model, hosted locally via Ollama, exhibited significantly higher response times (refer to Fig.~\ref{fig:llm_latency}). This illustrates the tradeoff between hosting models locally and using commercial APIs. 
   \item \textbf{Cost efficiency:} Token usage and costs vary widely. As shown in Table~\ref{tab:llm_benchmarking_matrix}, GPT-4o-mini offers the lowest cost per run, while GPT-5, followed by Claude 3.7, is substantially more expensive. These numbers reflect raw token costs for input/output and does not include expenses for fine-tuning or specialized hosting. 
   \item \textbf{OMNeT++ Configuration Accuracy:} The OMNeT++ configuration files generated by the models should ideally be ready for direct use in the simulator without any manual intervention. However, the off-the-shelf models tested with the given input prompt did not generate complete, simulation-ready configuration files (refer to Table~\ref{tab:llm_benchmarking_matrix}). Models classified as \textbf{Partially Correct} produced the \textit{.ned} and \textit{.ini} files, but the syntax was partially correct or several parameters were missing. Models that generated only a few lines of the \textit{.ned} and \textit{.ini} files, or hallucinated entirely irrelevant text, were classified as \textbf{Incorrect}. Interestingly, Claude 3.7 did not produce any OMNeT++ files and was therefore classified as \textbf{Not Attempted}.
   \item \textbf{Output Completeness:} The quality of generated outputs varied significantly across models (refer to Table~\ref{tab:llm_benchmarking_matrix}). Output quality was judged based on the ability of the model to produce complete responses for a given prompt. In general, most models produced syntactically coherent configurations that were aligned with the task description. However, certain models, such as Claude 3.7 and LLaMA 3.1-70B, often generated incomplete responses, with outputs terminating abruptly. 
   \item \textbf{Traffic Type Identification:} 
    Models such as GPT-5, Gemini 2.5 Pro, and Grok 4 assigned all flows as TT flows. However, models such as GPT-4o, GPT-4o-mini, and Claude 3.7 did not provide any traffic type assignment (refer to Table~\ref{tab:llm_benchmarking_matrix}). Interestingly, LLaMA 3.1-70B assigned priorities to the flows but did not specify the traffic type. Overall, the lack of explicit traffic type assignments in models limits their usefulness for TSN orchestration.
\end{itemize} 

\subsection{Takeaway}
This case study illustrates that current off-the-shelf LLMs are capable of parsing natural-language TSN orchestration tasks, however, the outputs reveal several key limitations, including substantial differences in variability in output quality, incorrect or incomplete OMNeT++ configuration files, and hallucinated traffic type assignments. \textit{Prompt design:} The effectiveness of an LLM in performing a task depends heavily on how the task is described in the prompt~\cite{survey_llm}. Clear and precise instructions are therefore essential for efficient performance. A critical takeaway is the need for structured prompt templates that effectively guide the LLM in TSN orchestration tasks. Future research should systematically evaluate prompt design and its impact on output accuracy. \textit{TSN benchmark:} Comprehensive evaluation and benchmarking of different LLMs on TSN tasks are necessary. This will clarify which models perform best and are suitable for fine-tuning. It is highly unlikely that all stages of TSN orchestration can be solved by LLMs. Furthermore, some models may excel in specific stages but fail in others. Selecting the best model for each stage therefore requires detailed comparative analysis. Finding the right TSN stage to evaluate is another research direction. \textit{External tools:} The results clearly show that none of the models perform adequately in traffic type and priority assignment. Thus, integrating external tools such as NC and TTA solvers~\cite{rubi_icc} remains essential. 

\section{Lessons Learned \& Open Research Directions}
\label{sec:research_direction}
\subsection{Benefits of LLM-aided TSN Orchestrator}
Configuring TSN networks directly through natural language, without requiring specialized modeling languages such as YANG, RestConf, XML, or JSON, makes LLM-based orchestration highly attractive for industry and real-world deployments. The main benefits include:

\begin{itemize}
    \item \textbf{Reduced engineering effort:} LLMs may reduce the time spent on repetitive configuration, translation, and documentation tasks across the TSN workflow, such as GCL configuration.
    \item \textbf{Assistance for complex workflows:} LLMs can help prepare inputs, such as converting the inputs from one format to another format, summarize outputs, and coordinate validation steps for complex TSN configurations. However, the final verification should remain with established state-of-the-art tools and engineering review by domain experts.
    \item \textbf{User-friendly system:} Non-experts can design deterministic TSN networks through natural language interfaces, making the system more accessible.  
\end{itemize}

\subsection{Challenges and Open Issues}
Despite these advantages, several challenges remain for realizing an LLM-aided TSN orchestrator:
\begin{itemize}
    \item \textbf{Mathematical module:} TSN WCD analysis relies on mathematical tools such as NC, which LLMs cannot perform reliably. Therefore, integrating dedicated computational modules and optimization solvers, as suggested in~\cite{ChatNet}, is necessary. Integrating LLMs with existing NC tools will provide the necessary mathematical guarantees for WCD analysis.
    \item \textbf{TSN datasets:} The absence of open-source datasets and algorithms limits the ability to fine-tune and verify models for TSN-specific tasks. Having open-source TSN test cases for different applications supports reproducibility.
    \item \textbf{Continuous verification:} Misconfigurations in TSN can cause wasted resources, QoS violations, or safety-critical failures. LLM outputs therefore require continuous verification and validation, with the help of NC solvers, such as the ones provided in~\cite{luxi_access, luxi_tnsm}, and existing TSN configuration tools, ideally with minimal human intervention.
    \item \textbf{Assurance and trust:} Current off-the-shelf models are not yet sufficient to replace the existing mathematical tools in assurance-relevant stages (C-E). For these stages, LLMs can assist in input preparation, output translation, report generation, and assistance of established tools. Future LLM-assisted TSN systems should move from one-shot generation toward hybrid workflows that combine LLMs with solvers, simulators, validators, and, where relevant, model checkers or theorem provers.
\end{itemize}

\vspace{-0.2cm}
\subsection{Integration with External Tools}
Hybrid integration with established tools is essential for practical deployment. 
Since LLMs perform poorly in mathematical reasoning~\cite{ChatNet}, tasks such as GCL generation (D), network configuration (C-D), schedulability analysis (E), WCD calculation (E), and TTA (C) should be handled or verified by external tools. Here we propose using NC tools for (E), such as those proposed in \cite{luxi_tnsm, luxi_access}, heuristics and mathematical algorithms to generate the GCL for (D), such as \cite{gcl_silviu, window_niklas, rubi_iotj}, metaheuristics and deep reinforcement learning (DRL)-based solutions for TTA, as proposed by \cite{voica_traffic_assignment, rubi_icc}, and OMNeT++ and NS-3 based simulation frameworks for (F-G), as provided by INET and \cite{rubi_rtcsa, rubi_noms, rubi_ccnc}.

\section{Conclusion and Future Directions}
\label{sec:future_work}
In this paper, we present a vision for Large Language Model (LLM)-assisted orchestration of Time-Sensitive Networking (TSN). We outline an end-to-end workflow that includes input parsing, traffic type assignment, scheduling and configuration, schedulability analysis, OMNeT++ file generation, simulation-based validation, log analysis, and monitoring. Unlike general network automation, TSN imposes deterministic guarantees and mixed-criticality constraints, making orchestration uniquely challenging. We argue that LLMs can serve as an automation layer when complemented by external tools such as NC engines and advanced scheduling algorithms. We also showcase illustrative case studies that examine the capabilities of existing off-the-shelf models. The results suggest that current off-the-shelf LLMs can already assist with selected TSN tasks, especially parsing structured prompts and generating preliminary configuration artifacts, though with clear tradeoffs in latency, cost, and output quality. However, these outputs are not yet sufficient as standalone assurance evidence. A more promising direction is hybrid orchestration, in which LLMs prepare inputs, coordinate tools, and summarize results, while external configuration, schedulability, simulation, and formal verification tools provide the evidence used for timing and deployment decisions. Looking ahead, we identify four research directions: (i)~developing open TSN datasets and standardized benchmarks for checkable TSN tasks, (ii)~fine-tuning or adapting LLMs for TSN-specific artifacts and workflows, (iii)~building hybrid orchestration systems that integrate LLMs with external configuration, schedulability, simulation, and formal verification tools, and (iv)~defining explicit assurance arguments for how LLM-generated artifacts are validated before deployment.

\bibliographystyle{IEEEtran}
\renewcommand{\baselinestretch}{0.87}
\bibliography{reference}
\end{document}